\documentclass[twoside,12pt]{article}
\usepackage[dvips]{graphics,epsfig}
\usepackage[latin1]{inputenc}
\usepackage{setspace}
\usepackage{textcomp}
\usepackage{amssymb,amsfonts,amsmath}
\usepackage{graphicx}
\usepackage{epstopdf}
\usepackage{subfig}

\def\laq{\raise 0.4 ex \hbox{$<$}\kern -0.8 em\lower 0.62 ex\hbox{$\sim$}}
\def\gaq{\raise 0.4 ex \hbox{$>$}\kern -0.7 em\lower 0.62 ex\hbox{$\sim$}}

\def\JP{{\it J. Phys.} }

\def\PR{{\it Phys. Rev.} }

\def\RMP{{\it Rev. Mod. Phys.} }

\def\beq{\begin{equation}}
\def\eeq{\end{equation}}
\def\bea{\begin{eqnarray}}
\def\eea{\end{eqnarray}}
\def\nn{\nonumber}

\def\r{\mathbf{r}}
\def\Re{\textrm{Re}}

\def\x{\mathbf{x}}

\def\p{\mathbf{p}}

\begin{document}
\pagestyle{plain}


\begin{center}

{\Large\bf Disordered Bose-Einstein condensate in hard walls trap}

\vspace*{1.0cm}

R. Acosta-Diaz$^{*}$, C. A. D. Zarro$^{**}$\\
\vspace*{0.5cm}
{Instituto de F\'isica, \\
Universidade Federal do Rio de Janeiro,\\
21941-972, Rio de Janeiro, RJ, Brazil }\\

\vspace*{1.0cm}

G. Krein$^{\circ}$\\
\vspace*{0.5cm}
{Instituto de F\'{i}sica Te\'orica, Universidade Estadual Paulista,\\
Rua Dr. Bento Teobaldo Ferraz, 271 - Bloco II,  01140-070 S\~ao Paulo, SP, Brazil }\\

\vspace*{1.0cm}

A. Saldivar$^{\dagger}$ and N. F. Svaiter$^{\dagger\dagger}$\\
\vspace*{0.5cm}
{Centro Brasileiro de Pesquisas F\'{\i}sicas,\\
Rua Dr. Xavier Sigaud 150,
22290-180, Rio de Janeiro, RJ, Brazil}\\

\vspace*{2.0cm}
\end{center}

\begin{abstract}
We discuss the effects of quenched disorder in a dilute
Bose-Einstein condensate confined in a hard walls trap.
Starting from the disordered Gross-Pitaevskii functional, we obtain a representation 
for the quenched free energy as a series of integer moments of the partition function. 
Positive and negative disorder-dependent effective coupling constants appear in the integer moments. 
Going beyond the mean-field approximation, we compute the static two-point correlation functions 
at first-order in the positive effective coupling constants. 
We obtain the combined
contributions of effects due to boundary conditions and disorder in this weakly disordered condensate. The ground state renormalized density 
profile of the condensate is presented. We also discuss the appearance of metastable
and true ground states for strong disorder, when the effective coupling constants become negative.
\noindent

\end{abstract}

\vfill
\noindent\underline{\hskip 140pt}\\[4pt]
{$^{*}$ E-mail address: racosta@if.ufrj.br} \\
\noindent
{$^{**}$ E-mail address: carlos.zarro@if.ufrj.br}\\
\noindent
{$^{\circ}$ E-mail address: gastao.krein@unesp.br}\\
\noindent
{$^{\dagger}$ E-mail address: asaldivar@cbpf.br}\\
\noindent
{$^{\dagger\dagger}$ E-mail address: nfuxsvai@cbpf.br}\

\newpage


\doublespacing

\section{Introduction}\label{intro}

Effects due to boundary conditions in the dilute gases and condensates have been 
discussed by many authors~\cite{walls,bec2,bec3}.  In particular, the influence of hard walls type 
of boundary conditions on the statistical properties of an ideal Bose-Einstein condensate was  
discussed in Ref.~\cite{scully}. In Ref.~\cite{kneer} the dynamics of an expanding condensate 
in a hard walls trap was investigated, finding that confinement generates structures in the 
condensate. Many other interesting situations can be created by coupling in a controlled 
manner disorder fields to a trapped Bose-Einstein condensate, allowing studies of the combined effect of boundary 
conditions and disorder.

Controlled disorder can be realized in ultracold Bose gases by using $e.g.$ laser speckle fields or 
incommensurate optical lattices \cite{lye,pelster}. Under the influence of disorder, a bosonic system 
at low temperatures can contain many fragmented condensates~\cite{noz,huang,baym,nattermann}. 
 There are two main theoretical methods to investigate the effects of disorder 
in the Bose-Einstein condensate; methods based on Bogolibov theory~\cite{bogoliubov,abrikosov},
and the replica method~\cite{re,dotsenko,parisi88}. For low temperatures, small interaction strength
and weak disorder, it is possible to use a perturbative approach.  For strong disorder, the effects 
of the disorder must be treated non-perturbatively~\cite{halperin,fisher,yukalov,yukalova}. To discuss the condensate in disordered environment, the 
replica method was used in Refs.~\cite{lopatin,zohay,axel}. 

Despite the  successes in the application of the replica method in 
disordered systems, a mathematical rigorous derivation to support this procedure is still lacking~\cite{cri1,cri2,cri4,cri5}. Intended to fill this gap, a mathematically 
rigorous technique for computing the average free energy of a system with quenched disorder in continuous field theory models was proposed  in 
Refs.~\cite{distributional,distributional2,zarro,zarro2}. The advantage of this approach is that contrary to the replica method, this technique 
neither involves derivatives of the $k$-th integer moments of the partition function 
$\mathbb{E}\left[Z^{k}\right]$ with respect to $k$, nor extension of these derivatives to non-integers 
values of $k$.

The purpose of this paper is to discuss the effects of disorder fields in a hard walls trapped dilute Bose-Einstein 
condensate.  Here we are
interested  in the situation of a quenched disorder field coupled to the particle density 
of the system. Using mainly functional methods~\cite{id, zinnjustin,lundh,anderson}, and with this 
new mathematical tool in hands,  we investigate the density profile of the low temperature trapped dilute-gas 
Bose-Einstein condensate for weak disorder fields.
Starting from a Gross-Pitaevskii energy functional, we employ the recently proposed
method to evaluate the quenched free energy of disordered 
systems for arbitrary strengths of the disorder.
A representation for the quenched 
free energy, described by a series of integer 
moments of the 
partition function is obtained.  This representation is non-trivial and very rich in consequences. 
For some critical value of the strength of the disorder, an 
effective negative coupling constant appears in some integer moments. We call this situation the strong disorder limit of the model.
Going beyond the 
mean-field approximation, we discuss the combined 
effect of the two-body interaction and a hard walls trap in the condensate for
the situation of positive effective coupling constants, i.e., the weak disorder situation. From the static two-point correlation functions in the one-loop approximation, we obtain the combined
contributions of effects due to boundary conditions and disorder on the 
density profile of the condensate. The 
renormalized density of the condensate which defines the ground state solution of the system is presented. To the best of our knowledge, 
these results are new. In addition, this is the first calculation of boundary effects 
in disordered Bose-Einstein condensates using field theoretical methods.
For the cases in which effective 
negative coupling constants appear in some integer moments, there are metastable states and true ground states \cite{stoof2,kagan,shuryak,g1,kagan2}. This 
is the scenario for first-order phase transitions induced by strong disorder fields. 

This paper is organized as follows. In Sec. \ref{sec:gpenergyfunctional} we discuss the disordered Gross-Pitaevskii energy functional. In Sec. \ref{sec:zeta} we derive the disorder average free energy of the model for arbitrary strengths of the disorder. In Sec. \ref{sec:effectiveenergy} we discuss the effective Gross-Pitaevskii functional. In Sec. \ref{sec:beyonda} we study the one-loop static two-point correlation functions in the presence of hard walls, in a generic integer moment of the partition function for the case where the  effective coupling constant is positive. In Sec. \ref{sec:sstatic} in this weak disorder limit, we obtain an expressions for the terms that contribute to the renormalized density  of the condensate. In Sec. \ref{sec:finalresult} the
renormalized density  of the condensate confined in a hard walls trap is presented. Conclusions are given in Sec. \ref{sec:conclusions}. In Appendix \ref{ap:firstorder} we present some detailed calculations which were omitted through Sec. \ref{sec:sstatic}. To simplify the calculations we assume the units to be such that $\hbar=c=k_{B}=1$.


\section{Gross-Pitaevskii energy functional}\label{sec:gpenergyfunctional}

We consider a system of identical nonrelativistic spinless bosons in a $d$-dimensional spatial volume, where we assume that $d\geq 3$. For dilute systems, one can consider the bosons structureless and use the framework of nonrelativistic field theory. In the Heisenberg representation, the bosonic field operators $\psi^{\dagger}(t,\x)$ and $\psi(t,\x)$ satisfy the usual equal-time canonical commutation relations:
\bea
&& [\psi^{\dagger}(t,\x),\psi^{\dagger}(t,\x')]=[\psi(t,\x),\psi(t,\x')]=0, \\
&& [\psi(t,\x),\psi^{\dagger}(t,\x')]=\delta^{d}(\x-\x').
\label{comm-rel}
\eea
For  interacting Bose particles, the simplest situation is the one with a mutual two-body interaction represented by a instantaneous two-body potential $V_{0}(|\x-\x'|)$, and an external potential $U(\x)$. In this case, the Hamiltonian operator for the system  is given by
\begin{eqnarray}
H(\psi^{\dagger},\psi) &=& \int d^{d}x \, \psi^\dag(t,\x) \left[-\frac{\Delta}{2m} 
+ U(\x)\right]\psi(t,\x) \nonumber \\
&+& \int d^{d}x d^{d}x'\,\psi^{\dagger}(t,\x)\psi^{\dagger}(t,{\x}')
V_{0}(|\x-\x'|)\psi(t,\x')\psi(t,{\x}),
\label{eq:hamiltonian}
\end{eqnarray}
where $m$ is the mass of the bosonic particles and $\Delta$ is the Laplacian in $\mathbb{R}^{d}$. The operator for the total number of particles of the system is defined as
\begin{equation}\label{eq:hamiltonian0}
N_{0}(\psi^{\dagger},\psi)=\int\,d^{d}x\,\psi^{\dagger}(\x)\psi({\x}).
\end{equation}
In the absence of the external potential $U(\x)$ the Hamiltonian operator is invariant under translations $(\x'=\x+\mathbf{c})$ and  SO(3) rotations in the three dimensional space. Also the Hamiltonian operator is invariant under a global rotation $\psi(\x)\rightarrow\psi'(\x) = e^{i\alpha}\psi(\x)$, for constant $\alpha$. The generator of the $U(1)$ gauge transformation is the particle number operator since $e^{-i\alpha\,N_{0}}\psi(\x)\,e^{i\alpha\,N_{0}} = \psi(\x)e^{i\alpha}$. The familiar Bogoliubov spectrum is the result of the spontaneous breaking of this continuous symmetry, a manifestation of the Goldstone theorem.

For dilute systems, one can replace the potential $V_{0}(|\x-\x'|)$ by a contact interaction. In this case, from Eq.~(\ref{eq:hamiltonian}), one has that the Heisenberg equation of motion for the field operator $\psi(t,\x)$ reads
\begin{equation}
i\frac{\partial}{\partial\,t}\psi(t,\x) = \bigl[\psi(t,\x),H(\psi^{\dagger},\psi)\bigr]= \left[-\frac{\Delta}{2m}+U(\x) +\frac{g}{4}|\psi(t,\x)|^{2}\right]\psi(t,\x),
\label{eq:timedependenta}
\end{equation}
where $g$ is the strength of the contact interaction, related to the scattering length $a$ by $g={16\pi a}/{m}$. To study the stationary solutions of this Hamiltonian, it is usual to decompose the field operator into a condensate part, represented by a classical field $\varphi(\x)$, the scalar order parameter, and a quantum fluctuating part, represented by ${\vartheta}(t,\x)$ as:
\beq
\psi(t,\x) = \left[\varphi(\x) + {\vartheta}(t,\x)\right]\,e^{- i\mu t}
\label{decomp}
\eeq
where $\mu$ is the chemical potential, which is determined by Eq. (\ref{eq:hamiltonian0}). When replacing this into Eq.~(\ref{eq:timedependenta}) under the hypothesis that the effects of ${\vartheta}(t,\x)$ are small, one obtains the time-independent Gross-Pitaevskii equation~\cite{gross,pit,gross2},
\begin{equation}
\left[-\frac{\Delta}{2m}+U(\x)-\mu +\frac{g}{4}|\varphi(\x)|^{2}\right]\varphi(\x)=0.
\end{equation}
As it is well known, this hypothesis is valid under certain conditions, namely, for a large number of  particles and when the $s$-wave scattering length is much smaller than the mean separation between the particles. When treating the term ${\vartheta}(t,\x)$ to first order in perturbation theory, one obtains the Bogoliubov spectrum of the quantum excitations of the condensate. 

In many experimental setups the disorder couples to the particle density. In that case, one can write $U(\x) = V_{\text{trap}}(\x) + V(\x)$, where  $V_{\text{trap}}(\x)$ is the potential representing the trap and $V(\x)$ represents the disorder potential. In the following discussions, the shape of the trap potential is not important. We consider the Gross-Pitaesvikii framework, i.e. we neglect the fluctuating part of the field operator. For time-independent $V_{\text{trap}}$ and quenched disorder, we can employ the formalism of statistical field theory. 

We remark that there is another approach to obtain a classical field theory.  Suppose that we write the grand-canonical partition function as a path integral over complex field $\psi(\tau,\x)$ and $\psi^{*}(\tau,\x)$. Finite temperature effects are introduced where the energy integration is replaced by sum over the Matsubara frequencies. Therefore the complex field is a periodic function of the imaginary time with period $\beta$.  After expand the field $\psi(\tau,\x)$ and $\psi^{*}(\tau,\x)$ in the path integral in terms of the Fourier components and restrict to the zero Matsubara component, the fields $\psi$ and $\psi^{*}$ can be considered as classical fields. The partition function is written in terms of a classical field theory where $\varphi_{0}(\x)$ depends only on the spatial coordinates. This is called a classical field approximation.

Going back to our problem, the central object of interest is the disordered average free energy of the system, that is  obtained from the logarithm of the disordered partition function:
\begin{equation}
\label{eq:partition}
Z(V)=\int[d\varphi][d\varphi^{*}]\exp\left[-\beta E(\varphi^{*},\varphi, V\bigr)\right],
\end{equation}
where $[d\varphi][d\varphi^{*}]$ is a measure in the space of all order-parameter configurations, with the disordered Gross-Pitaevskii energy functional $E(\varphi^{*},\varphi, V)$ given by
\begin{equation}
\label{eq:energyfunctional}
E = \int d^{d}x\, \left[\varphi^{*}(\x) \left(-\frac{\Delta}{2m} + V_{\text{trap}}(\x)
+ V(\x)-\mu\right)\varphi(\x)  + \, \frac{g}{4}\bigl(\varphi^{*}(\x)\varphi(\x)\bigr)^{2}\right].
\end{equation}
To obtain the quenched free energy, $i.e.$ the average over the ensemble off all realizations of the quenched disorder, one needs to provide a probability distribution for the random 
potential $V(\x)$, which we assumed as a delta-correlated Gaussian distribution with strenght $\sigma$: 
\begin{equation}
\label{eq:deltacorrelated}
\mathbb{E}[V(\x)]=0\;\;\;\;\text{ and }\;\;\;\mathbb{E}[V(\x)V(\x')] = \sigma \, \delta^{d}(\x-\x').
\end{equation}

In the next section, we briefly review the method based on the distributional zeta function \cite{distributional, distributional2,zarro,zarro2} to compute the quenched free energy.


\section{The quenched free energy}
\label{sec:zeta}

Given the probability measure, we are able to  compute averages of physical quantities. Our main interest is to compute the quenched free energy:
\begin{equation}
F=-\frac{1}{\beta}\int [d\,V]P(V)\,\ln Z(V),
\label{pro1}
\end{equation}
where  $[d\,V]$ is a functional measure. Connected correlation functions are obtained by taking derivatives of $F$ with respect to an external complex source coupled to the complex order parameter in Eq. \eqref{eq:energyfunctional}, as usual.

Inspired by situations in which one defines zeta functions in terms of countable collection of numbers, as for example prime numbers, non-trivial zeros of the Riemann zeta function, among others, we define the distributional zeta-function $\Phi(s)$  as
\begin{equation}
\Phi(s)=\int [d\,V]P(V)\,\frac{1}{Z(V)^{s}},\label{pro1}
\end{equation}
\noindent for $s\in \mathbb{C}$, this function being well defined in an open connected subset of the complex plane. The quenched free energy is obtained from Eq. \eqref{pro1} as
\begin{equation}
F=\frac{1}{\beta}\left(\frac{d}{ds}\right)\Phi(s)|_{s=0^{+}}, \,\,\,\,\,\,\,\,\,\, \Re(s) \geq 0,
\end{equation}

Next, use of  the Euler's integral representation for the gamma function $\Gamma(s)$ \cite{rizik} allows us to write $1/Z(V)^{s}$ as
\begin{equation}
\frac{1}{Z(V)^{s}}=\frac{1}{\Gamma(s)}\int_{0}^{\infty}dt\,t^{s-1}e^{-Z(V)t}.
\end{equation}
Breaking this integral into two integrals, one from $0$ to $a$ and another from $a$ to $\infty$ where $a$ is an arbitrary dimensionless real number, and expanding the exponential in powers of $t$ in the first integral, one can write  the quenched free energy  as: 
\begin{equation}\label{eq:completefreeenergy}
F=\frac{1}{\beta}\Bigg[\sum_{k=1}^{\infty} \frac{(-1)^{k}a^{k}}{kk!}\mathbb{E}\left[Z^{k}\right] +\log{a} + \gamma - R(a)\Bigg],
\end{equation}
where $\mathbb{E}\left[Z^{k}(V)\right]$ is the $k$-th moment of the partition function
\begin{equation}\label{eq:EZk}
\mathbb{E}\left[Z^{k}(V)\right]=\int [d\,V]P(V)\,Z^{k}(V),
\end{equation}
$\gamma$ is the Euler constant and $R(a)$ is given by
\begin{equation}
R(a)=\int [dV]P(V)\int_{a}^{\infty}\,\dfrac{dt}{t}\, e^{-Z(V)t}.
\end{equation}
Notice that the disorder average free energy defined by Eq. (\ref{eq:completefreeenergy}) is $a$ independent.

The crucial question is the physical interpretation of Eq. \eqref{eq:completefreeenergy}.  This equation indicates that the disordered system may be represented by an ensemble of subsystems, each of them being described by an integer moment of the partition function. In systems described by infinitely many degrees of freedom  without disorder, the fundamental object is the generating functional of connected correlation functions, which are the moments of a probability measure. To describe disordered systems, the disordered-average generating functional of connected correlation functions is written as a series of the moments of the generating functional of correlation functions. As the result of such generalization, any renormalized quantity, as for example, the renormalized density of the condensate will be written also in a series of contributions coming from all moments of the partition function. This issue will be clarified further ahead in this paper.

Our next goal is to examine the fields in each integer moment of the partition function, for the concrete case of interest, the Gross-Pitaevskii energy functional of Eq.\eqref{eq:energyfunctional}. This will be done in the next section.


\section{The effective Gross-Pitaevskii energy functional} \label{sec:effectiveenergy}

We have now at our disposal the Eq. \eqref{eq:completefreeenergy}, which, for large $a$, can be written as
\begin{equation}\label{eq:incompletefreeenergy}
 F=\frac{1}{\beta}\sum_{k=1}^{\infty} \frac{(-1)^{k}}{kk!}\mathbb{E}\left[Z^{k}\right],
\end{equation}
where $a^{k}$ factors have been absorbed in the functional measure of each integer moment of the partition function by redefining the fields which then amounts to redefining the parameters of the energy functional.

The object of interest is the $k$-th integer moment of the partition function, $\mathbb{E}\left[Z^{k}\right]$. From Eqs.~\eqref{eq:partition},~\eqref{eq:energyfunctional},~\eqref{eq:deltacorrelated} and Eq.~\eqref{eq:EZk} one can show that
\begin{equation}\label{eq:ezk0}
\mathbb{E}\left[Z^{k}\right]=\int \prod_{i=1}^{k} \left[ d\varphi_{i}^{*(k)}\right]\left[ d\varphi_{i}^{(k)}\right]
e^{-\beta E_{\text{eff}}^{(k)}\left(\varphi_{i}^{*(k)},\varphi_{i}^{(k)}\right)},
\end{equation}
where 
\begin{equation}
\label{eq:effectivehamiltonian}
E_{\text{eff}}^{(k)}(\varphi_{i}^{*(k)},\varphi_{i}^{(k)})= \int d^{d}x  \sum_{i=1}^{k}  
\varphi_{i}^{*(k)}(\x) \Biggl[ -\frac{\Delta}{2m}+V_{\text{trap}}+ \mu +  \frac{1}{4}\sum_{j=1}^{k}g_{ij}|\varphi_{j}^{(k)}(\x)|^{2}\Biggr]\varphi_{i}^{(k)}(\x),
\end{equation}
\noindent
and the generalized coupling constant $g_{ij}$ defined by
\beq
g_{ij} = g\delta_{ij} - \beta\sigma.
\eeq
For simplicity, we write $V_{\text{trap}}(\x)=V_{\text{trap}}$.
The saddle-point equations derived from the $k$-th moment  read 
\begin{equation}
\label{eq:spe}
\left(-\frac{\triangle}{2m}+V_{\text{trap}}-\mu + \sum_{j=1}^{k} \frac{g_{ij}}{2} |\varphi_{j}^{(k)}(\x)|^{2}\right) 
\varphi_{i}^{(k)}(\x)=0.
\end{equation}
Using that $\varphi_{i}^{(k)}(\x)=\varphi_{j}^{(k)}(\x)$, the above equation becomes
\begin{equation}\label{eq:repsymspe}
\left(-\frac{\triangle}{2m}+V_{\text{trap}}-\mu + \frac{\lambda_{k}}{2}|\varphi_{i}^{(k)}(\x)|^{2}\right)\varphi_{i}^{(k)}(\x)=0,
\end{equation}
where we have defined the effective coupling constant $\lambda_{k}=g-\beta\sigma k$. 

Consider the series in Eq. (\ref{eq:incompletefreeenergy}) where integer moment of the partition function is given by $\mathbb{E}\,[Z^{\,k}]$ {\textemdash} see also Eqs. (\ref{eq:ezk0}) and (\ref{eq:effectivehamiltonian}).  In our formalism, we have introduced auxiliary fields to describe the disordered condensate. As we discussed before, the consequence of such approach is a totally new picture for disordered systems where the ensemble of subsystems are not individually observed. Now we can imagine that there is a mapping between the number of particles of the condensate $n_{0}$, and the number of auxiliary fields. Since the number of bosonic particles in the condensate is finite, the number of fields also must be finite. We are claiming that the truncation of the series does not affect our ability to describe the disordered condensate. The consequence of such choice, is that the series given by Eq. \eqref{eq:incompletefreeenergy} is truncated for some $N$. Hence, the structure of the  fields in each integer moment of the partition  function can be defined as
\begin{equation}
\begin{cases}
\varphi_{i}^{(k)}(\x)=\varphi^{(k)}(\x) \,\,\,\hfill\hbox{for $k=1,2,...,N$}\\
\varphi_{i}^{(k)}(\x)=0 \quad \,\,\,\,\,\hbox{for $k>N$}.
\end{cases} \label{RSB1}
\end{equation}
This leads to the effective Gross-Pitaevskii energy functional:
\begin{equation}\label{eq:effectivehamiltonian2}
E_{\text{eff}}^{(k)}= k\int d^{d}x \left[\varphi^{*(k)}(\x)\Bigl(-\frac{\Delta}{2m}
+ V_{\text{trap}}-\mu \Bigr)\varphi^{(k)}(\x) + \frac{\lambda_k}{4}  \varphi^{*(k)}(\x)|\varphi^{(k)}(\x)|^{2}
\varphi^{(k)}(\x)\right].
\end{equation}

Due to the disorder, to describe the system we have to take into account contributions from fields  governed by different effective coupling constants $\lambda_{k}$.  It is important to stress that from these effective coupling constants it is possible to find the  renormalized boson-boson interactions in the one-loop approximation using all the terms of the series for the average free energy. This issue will be investigated in the next sections where we will  obtain the condensate density from the two-point correlations functions of the model.

For fixed $N$, let us define the weak disorder limit as the case when the effective coupling constants $\lambda_{k}$ in all $k$-th moments are positive. In the more general situation, there are two sets of moments: the integer moments where effective coupling constants are positive and a second set where the effective coupling constants become negative for some critical value for the strength of disorder. We called this situation the strong disorder limit of the model. Let us assume that this happens for some critical $k_{c}$. In the series representation for the quenched free energy there is a critical $k_{c}=\lfloor \frac{g}{\beta\sigma}\rfloor$. In this case the quenched free energy can be written as 
\begin{equation}\label{eq:completefreeenergyseries}
F=\frac{1}{\beta}\sum_{k=1}^{k_{c}}\frac{(-1)^{k}}{k!k}\mathbb{E}[Z^k]+\frac{1}{\beta}\sum_{k=k_{c}+1}^{N}\frac{(-1)^{k}}{k!k}\mathbb{E}[Z^k].
\end{equation}

Let us investigate a generic integer moment of the series  with $k < k_c$. Defining $\chi = \varphi^{(k)}$, we have
\begin{equation}\label{eq:ezk2}
\mathbb{E}\left[Z^{k}\right] =  {\cal N} \int\left[ d\chi\right] \left[ d\chi^{*}\right]
e^{-\beta E_{\text{eff}}^{(k)}\left(\chi,\,
\chi^{*}\,\right)},
\end{equation}
where $\left[ d\chi\right] \left[ d\chi^{*}\right]$ is a functional measure and $\cal{N}$ is a normalization factor, and
\begin{equation}
\label{eq:effectivehamiltonian22}
 E_{\text{eff}}^{(k)}(\chi,\chi^{*}) = k \int d^{d}x\, \left[\chi^{*}(\x)\Bigl(-\frac{\Delta}{2m}
+ V_{\text{trap}}-\mu \Bigr)\chi(\x)  \frac{\lambda_k }{4} \chi^{*}(\x) |\chi(\x)|^{2} \chi(\x)\right].
\end{equation}
We are interested  in calculating the ground states of this energy functional in the weak disorder limit. For accessing them, we work with the Cartesian representation for the complex field $\chi(\x)$. We define the real fields $\phi_{1}(\x)$  and $\phi_{2}(\x)$ such that
\begin{equation}\label{eq:ezk33}
\chi(\x)=\frac{1}{\sqrt{2}}\bigl[\phi_{1}(\x)+i\phi_{2}(\x)\bigr]
\end{equation}
and
\begin{equation}\label{eq:ezk333}
\chi^{*}(\x)=\frac{1}{\sqrt{2}}\bigl[\phi_{1}(\x)-i\phi_{2}(\x)\bigr].
\end{equation}
The effective energy functional for these two real fields can be written as
\begin{align}\label{eq:effectivehamiltoniantt}
 E^{(k)}_{\text{eff}}(\phi_{1},\phi_{2}) &=k\int d^{d}x\left[\frac{1}{2}\phi_{1}(\x)\Bigl(-\frac{\Delta}{2m}
+ V_{\text{trap}}-\mu \Bigr)\phi_{1}(\x)\right. \nonumber \\
&\left.+\frac{1}{2}\phi_{2}(\x)\Bigl(-\frac{\Delta}{2m}
+ V_{\text{trap}}-\mu \Bigr)\phi_{2}(\x)+\frac{\lambda_{k}}{16}\bigl(\phi_{1}^{2}(\x)+\phi_{2}^{2}(\x)\bigr)^{2}\right].
\end{align}
Let us define the functional potencial, $\Phi^{(k)}(\phi_{1},\phi_{2})$, which is bounded from bellow, as
\begin{equation}\label{eq:effectivehamiltonianaaa}
\Phi^{(k)}(\phi_{1},\phi_{2})=k\int d^{d}x\,\left[\frac{1}{2}(-\mu)\bigl(\phi_{1}^{2}(\x)+\phi_{2}^{2}(\x)\bigr)
+\frac{\lambda_{k}}{16}\bigl(\phi_{1}^{2}(\x)+\phi_{2}^{2}(\x)\bigr)^{2}\right].
\end{equation}
The $O(2)$ symmetry corresponds to  an invariance under rotations in the real order parameters $(\phi_{1},\phi_{2})$ plane. The extremum of the potential functional  in Eq.~(\ref{eq:effectivehamiltonianaaa}), for $\mu > 0$ and the effective coupling constants $\lambda_{k}>0$, is at $\bigl(\phi_{1}^{2}(\x)+\phi_{2}^{2}(\x)\bigr) = v^{2}$, where $v^{2} = 4 \mu/\lambda_k$. The points on the circle  with squared radius $v^{2}$ in the field $(\phi_{1},\phi_{2})$ plane are the minima of the potential.  There is an infinite number of  degenerate ground states. This is the standard situation  of spontaneous symmetry breaking. Defining   $\phi'_{1}(\x) = \phi_{1}(\x)-v$ and $\phi'_{2}(\x)=\phi_{2}(\x)$,  the new energy functional  $E_{\text{eff}}(\phi'_{1},\phi_{2})$ is given by 
\begin{align} \label{eq:effectivehamiltonian33}
E_{\text{eff}}(\phi'_{1},\phi_{2}) &= k\int d^{d}x\,\Biggl\{\frac{1}{2}\phi'_{1}(\x)
\left[-\frac{\Delta}{2m} + 2 \mu + V_{\rm trap}\right] \phi'_{1}(\x) + \frac{1}{2}\phi_{2}(\x) \left(-\frac{\Delta}{2m}+ V_{\rm trap}\right) \phi_{2}(\x) \nn \\
& + \frac{\lambda_k v }{4} \phi'_{1}(\x)
\bigl(\phi_{1}^{\prime 2}(\x)+\phi_{2}^{2}(\x)\bigr)  
+ \frac{\lambda_k}{16}\bigl(\phi^{\prime 2}_{1}(\x) +\phi_{2}^{2}(\x)\bigr)^{2}\Biggr\}.
\end{align}

Two points are worth stressing. First, observables describing a stationary  state in which the condensate structure is time independent can be expressed in terms of the static connected correlation functions. Second, 
in a standard field theory scenario, to go beyond the saddle-point approximation, one calculates the effective potential. This approach is suitable for systems  without boundaries and a different technique needs to be considered
in the present case of a trapped system, where translational invariance is broken. 

For instance, in a field theory scenario it is difficult to study systematically finite size effects. The geometry of the system and boundary conditions determine the correlation functions of the model. Our aim is to investigate the combined effect of a hard walls trap  and the  interactions on the density profile of the disorder condensate. The  density can be expressed in terms of the two-point static correlation functions and to go beyond the saddle-point approximation we consider the one-loop corrections to such a zeroth-order result. This approach is able to exhibit the 
ground state solution of the Bose-Einstein
condensate for weak disorder and the presence of the trap. The ground state solution for the trapped Bose-Einstein condensate without disorder, for repulsive boson-boson interaction in $d=2$ and $d=3$ was discussed in Ref. \cite{lieb}.

A further step is taken if we remember that the number of particles in the condensate is given in terms of the density $\rho(\x)$ as 
\begin{equation}
n_{0} = \int d^{d}\x\, \rho(\x) = - \frac{\partial}{\partial\mu} F.
\end{equation}
Defining a series representation for $\rho(\x)$ through 
\begin{equation}
\int d^{d}x\; \rho(\x) = \int d^{d}x\,\sum_{k=1}^{N} \rho^{(k)}(\x),
\label{aa1}
\end{equation}
one can write
\begin{equation}
\int d^{d}x\, \rho^{(k)}(\x) = \frac{1}{\beta}\frac{(-1)^{k+1}}{k!k}
\frac{\partial}{\partial\mu}\mathbb{E}\,[Z^{\,k}].
\label{aa3}
\end{equation}
Using Eqs. \eqref{eq:completefreeenergyseries}, \eqref{eq:ezk2} and \eqref{eq:effectivehamiltonian33}, we get
\begin{equation}\label{eq:rhok}
 \rho^{(k)}(\x)=
\frac{(-1)^{k}}{k!}m\,G_{1}^{(k)}(\x,\x),
\end{equation}
where $G_{1}^{(k)}(\x,\x)$ is the static two-point correlation function associated to the field $\phi'_{1}(\x)$ at coincident points, obtained from $k$-th integer moment of the partition function. Using dimensional analysis, the factor
$m$ in the above equation must be introduced.
In principle it is possible to use a resummation procedure \cite{dolan,adolfo,nami2}, summing certain classes of diagrams from all orders of perturbation theory. Here we are using the one-loop approximation. In the following we will discuss the zeroth-order contribution and first-order corrections in the effective coupling constants to derive the two-point correlation function associated to the field $\phi'_{1}(\x)$.


\section{The two-point static correlation functions in the presence of the trap} 
\label{sec:beyonda}

We consider a hard walls potential in a generic $d$-dimensional space with $d-1$ unbounded coordinates $\r = (x^{1},x^{2},\cdots,x^{d-1})$ and one bounded coordinate $ 0 \leq z \leq L$. Therefore we have $\x = (\r,z)$. The hard walls potential is defined by $V_{\text{trap}}=0$, for $0<z<L$ and $V_{\text{trap}}=\infty$ for $z\,\leq 0$ and $z\,\geq L$. We impose Dirichlet boundary conditions  on the fields at the planes $z=0$ and $z=L$:
\begin{equation}
\phi'_{1}(\r,z)|_{z=0}=\phi'_{1}(\r,z)|_{z=L}=0,
\label{dir}
\end{equation}
and
\begin{equation}
\phi_{2}(\r,z)|_{z=0}=\phi_{2}(\r,z)|_{z=L}=0.
\label{new}
\end{equation}

It is more convenient to use the notation $\varphi(\x)$ for both fields $\phi'_{1}(\x)$ and $\phi_{2}(\x)$. The fields that satisfy Dirichlet-Dirichlet boundary conditions on $z=0$ and $z=L$  can be expanded in Fourier series as:
\begin{equation}
\varphi(\r,z)=\frac{1}{(2\pi)^{\frac{d-1}{2}}}
\sum_{n=1}^{\infty} u_{n}(z)
\int
d^{d-1}p\, \phi_{n}(\p)\, e^{i\,\p.\r},
\label{field}
\end{equation}
where $\p$ is the continuum momentum and $u_n(z)$ stands for the eigenfunctions of the operator
$-\frac{d^2}{dz^{2}}$:
\begin{equation}
-\frac{d^{2}}{dz^{2}}u_{n}(z)=k^{2}_{n}u_{n}(z),
\label{ult}
\end{equation}
\space
satisfying the completeness and orthonormality relations. Due to the boundary condition we have $k_{n}=\frac{n\pi}{L}$, $n=1,2,3,...$.  

From Eq. \eqref{field}, the static two-point correlation functions associated  with 
the $\phi'_{1}(\x)$ field, $G_{1}^{(0)}(\r,z,\r^{\prime},z^\prime,\mu)$ can be expressed as:
\begin{equation}
G_{1}^{(0)}(\r,z,\r^{\prime},z^\prime,\mu)=
\frac{m}{(2\pi)^{d-1}}\sum_{n=1}^{\infty}
u_{n}(z)u_{n}^{*}(z')\int d^{d-1}p
\frac{e^{i\,\p.(\r-\r^{\prime})}}{(\p^{2}+\left(\frac{n\pi}{L}\right)^{2}+4\mu m)}.
\label{geex}
\end{equation}
The static two-point correlation functions associated  with 
the field $\phi_{2}(\x)$ is $G_{2}^{(0)}(\r,z,\r^{\prime},z^\prime) = G_{1}^{(0)}(\r,z,\r^{\prime},z^\prime,\mu)|_{\mu=0}$. 
Inspection of Eq. \eqref{eq:effectivehamiltonian33} reveals that there are three- and four-point vertices. Let us denote $G_{1}^{(1)}(\r_{1},z_{1},\r_{2},z_{2},\mu)$, the one-loop correction to $G_{1}^{(0)}(\r,z,\r^{\prime},z^\prime,\mu)$ of order $\lambda_{k}$ and   $G_{1}^{(2)}(\r_{1},z_{1},\r_{2},z_{2},\mu)$, for corrections of order $(\lambda_{k}v)^{2}$, which is actually of order $\lambda_{k}$. The one-loop correction to the two-point correlation function that we defined as $G_{1}^{(1)}(\r_{1},z_{1},\r_{2},z_{2},\mu)$ is given  by

\begin{align}
G_{1}^{(1)}(\r_{1},z_{1},\r_{2},z_{2},\mu) &=\frac{3m\,\lambda_{k}}{4}\,\int\,d^{d-1}r\!\int_{0}^{L}\!\!\!\!dz
\,G_{1}^{(0)}(\r_{1},z_{1},\r,z,\mu)\nonumber \\
&\times\biggl[3\,G_{1}^{(0)}(\r,z,\r,z,\mu)+G_{2}^{(0)}(\r,z,\r,z)\biggr]
G_{1}^{(0)}(\r,z,\r_{2},z_{2},\mu),
\label{MF}
\end{align}
The second one, given by $G_{1}^{(2)}(\r_{1},z_{1},\r_{2},z_{2},\mu)$ has two contributions. We can write that $G_{1}^{(2)}(\r_{1},z_{1},\r_{2},z_{2},\mu)=G_{1a}^{(2)}(\r_{1},z_{1},\r_{2},z_{2},\mu)+G_{1b}^{(2)}(\r_{1},z_{1},\r_{2},z_{2},\mu)$. These one-loop contributions are given respectively by
\begin{align}
G_{1a}^{(2)}(\r_{1},z_{1},\r_{2},z_{2},\mu)&=\left(\frac{m \lambda_{k}v}{2}\right)^{2}\int d^{d-1}r\int d^{d-1}r^{\prime}
\int_{0}^{L}dz \int_{0}^{L}dz^{\prime}\;G_{1}^{(0)}(\r_{1},z_{1},\r,z,\mu) \nonumber \\
&\times\left[\frac{9}{2}\left(\,G_{1}^{(0)}(\r,z,\r^{\prime},z^{\prime},\mu)\right)^2+\frac{1}{2}\left(G_{2}^{(0)}(\r,z,\r^{\prime},z^{\prime})\right)^2\right]
G_{1}^{(0)}(\r^{\prime},z^{\prime},\r_{2},z_{2},\mu)
\label{nova}
\end{align}
%
and 
\begin{align}
G_{1b}^{(2)}(\r_{1},z_{1},\r_{2},z_{2},\mu)&=\left(\frac{m \lambda_{k}v}{2}\right)^{2}\int d^{d-1}r\int d^{d-1}r^{\prime}
\int_{0}^{L}dz \int_{0}^{L}dz^{\prime}\;G_{1}^{(0)}(\r_{1},z_{1},\r,z,\mu) G_{1}^{(0)}(\r,z,\r^{\prime},z^{\prime},\mu)
\nonumber \\
&\times\left[\,3G_{1}^{(0)}(\r^{\prime},z^{\prime},\r^{\prime},z^{\prime},\mu)+3G_{2}^{(0)}(\r^{\prime},z^{\prime},\r^{\prime},z^{\prime})\right]
G_{1}^{(0)}(\r,z,\r_{2},z_{2},\mu).
\label{nova1}
\end{align}

In the next section we will obtain the correction to the density profile of the condensate due to the disorder and the presence of the hard walls trap in first-order approximation of the effective 
coupling constant.  A detailed study of these two-point correlation functions of the fields $\phi'_1(\x)$ and $\phi_2(\x)$ in the  context of a Euclidean field theory with boundaries can  be found in Refs.~\cite{fo,caicedo,robson,apariciomestrado}. Since we are taking the continuum limit of this field theory, all the ultraviolet divergences can be controlled using an analytic regularization procedure \cite {bolini} with a renormalization scheme. It provides an adequate description of the effective long-distance physics in scalar field theory models.


\section{The density profile of the condensate due to the disorder and hard walls trap}\label{sec:sstatic}

The aim of this section is to present, after a regularization and a renormalization procedure, the contribution to the density profile of the condensate coming from some integer moment of the partition function,  produced by the disorder and the presence of a hard walls trap. As discussed before, we study this correction in first-order approximation of the effective coupling constant in the weak disorder limit. From a generic $k$-th moment of the partition function and using Eqs. \eqref{eq:rhok}, \eqref{geex}, \eqref{MF} and \eqref{nova}, $\rho^{(k)}(z)$ is expressed as:  
\begin{equation}
\rho^{(k)}(z)=\frac{(-1)^{k}}{k!}\left(\Delta\rho^{(0)}(z,\mu)+\Delta\rho^{(1)}(z,\mu,\lambda_{k})+\Delta\rho^{(2)}(z,\mu,\lambda_{k}^{2})\right),
\label{mu}
\end{equation}
where $\Delta\rho^{(0)}(z,\mu)=\Delta\rho_{1}^{(0)}(z,\mu)$, $\Delta\rho^{(1)}(z,\mu,\lambda_{k})$ and $\Delta\rho^{(2)}(z,\mu,\lambda_{k})$ are the contributions to the density profile from the two-point static correlation functions. To proceed we write $\Delta\rho^{(1)}(z,\mu,\lambda_{k})$ and  $\Delta\rho^{(2)}(z,\mu,\lambda_{k})$ as
\begin{equation}
\Delta\rho^{(1)}(z,\mu,\lambda_{k})=\Delta\rho_{1}^{(1)}(z,\mu,\lambda_{k})+\Delta\rho_{12}^{(1)}(z,\mu,\lambda_{k})
\label{eq:DeltaRho1}
\end{equation}
and
\begin{equation}
\Delta\rho^{(2)}(z,\mu,\lambda_{k})=\Delta\rho_{1}^{(2)}(z,\mu,\lambda_{k})+
\Delta\rho_{12}^{(2)}(z,\mu,\lambda_{k})
\label{eq:DeltaRho2}
\end{equation}
respectively.  In the expressions above, the  superscripts on the $\Delta\rho$  denote the power of the effective coupling constant 
in each contribution, and the subscripts denote that such contributions come from only the field 
$\phi'_{1}(\x)$ or both fields $\phi'_{1}(\x)$ and $\phi_{2}(\x)$.

\subsection{Zeroth-order result to the density profile}\label{subsec:zeroth}

In order to get the first contribution to the density profile from the two-point static correlation functions, which is of zeroth-order in $\lambda_{k}$, let us define $\Delta\rho_{1}^{(0)}(z,\mu)$ as 
\begin{equation}
\Delta\rho_{1}^{(0)}(z,\mu)=mG_{1}^{(0)}(\r,z,\r,z,\mu).
\end{equation}
To compute this quantity, we introduce  $T(z,\mu)\equiv m^{-1}G_{1}^{(0)}(\r,z,\r,z,\mu)$ as 
\begin{equation}
T(z,\mu)=\frac{1}{(2\pi)^{d-1}}\sum_{n=1}^{\infty}
u_{n}^{2}(z) \int d^{d-1}p \frac{1}{(\p^{\,2}+(\frac{n\pi}{L})^{2}+4\mu m)}.
\label{TadDD}
\end{equation}
Using the fact that the mode solutions of Eq. \eqref{ult} are $u_{n}(z)=\sqrt{\frac{2}{L}}\sin\left(\frac{n\pi\,z}{L}\right)$ and the following identity:
\begin{equation}
\sum_{n=1}^{\infty} \frac{\cos nx}{n^{2}+a^{2}}=-\frac{1}{2a^{2}}+\frac{\pi}{2a}\frac{\cosh a(\pi-x)}{\sinh \pi a},
\end{equation}
the quantity $T(z,\mu)$ can be split into two contributions:
\begin{equation}
T(z,\mu)=f_1(\mu)-f_2(z,\mu),
\label{split-TDD}
\end{equation}
the first one, $f_{1}(\mu)$, is independent of the distance to the hard wall  and  $f_{2}(z,\mu)$ depends on that distance. They are given by
\begin{equation}
f_1(\mu)=
\frac{1}{2(2\pi)^{d-1}L}
\sum_{n=-\infty}^{\infty}
\int \frac{d^{d-1}p}
{(\p^{\,2}+(\frac{n\pi}{L})^{2}+4\mu m)}
\label{f1}
\end{equation}
and
\begin{equation}
f_{2}(z,\mu)=
\frac{1}{2(2\pi)^{d-1}}\int{d}^{d-1}p\frac{1}{\sqrt{\p^2+4\mu m}}
\frac{\cosh((L-2z)\sqrt{\p^2+4\mu m})}{\sinh(L\sqrt{\p^2+4\mu m})}.
\label{f2}
\end{equation}
In the infinite cut-off limit, it appears volume divergences in $f_1(\mu)$ and surface divergences in 
$f_{2}(z,\mu)$. Surface divergences are the ones that depend on the distance to the hard walls. We will discuss 
these divergences later on. Using the well known identity of dimensional 
regularization \cite{as,bo,to}, i.e.,
\begin{equation}
\int\frac{d^{d}k}{(k^{2}+a^2)^{s}}=\frac{\pi^{d/2}}{\Gamma(s)}\Gamma\Bigl(s-\frac{d}{2}\Bigr)\frac{1}{a^{2s-d}},
\label{mu2}
\end{equation}
we can write $f_1(\mu)$ as
\begin{equation}
f_1(\mu)=\frac{\Gamma\bigl(\frac{3-d}{2}\bigr)}{2^d\pi^{\frac{5-d}{2}}L^{d-2}}\sum_{n=-\infty}^{\infty}\frac{1}{(n^{2}+\frac{4\mu m L^{2}}{\pi^2})^{\frac{3-d}{2}}}.
\label{mu3}
\end{equation}
The regularization of the above equation can be done using the definition of the modified Epstein zeta function.
The modified Epstein zeta function $\xi(s,a)$ is a function of the complex variable $s=\sigma+i t$, where $\sigma, t\,\, \in{\mathbb{R}}$, for $a^{2}>0$ defined by the absolutely convergent series as
\begin{equation}
\xi(s,a)=\sum_{n=-\infty}^{\infty}\frac{1}{(n^{2}+a^{2})^{s}},\,\,\,\,Re(s)>\frac{1}{2}
\label{nana}
\end{equation}
and in the whole complex plane $\mathbb{C}$ by analytic continuation, which is given by
\begin{equation}
\xi(s,a)=\frac{\sqrt{\pi}}{a^{2s-1}\Gamma(s)}\bigg[\Gamma\left(s-\frac{1}{2}\right)+4\sum_{n=1}^{\infty}(\pi n a)^{s-1/2}K_{s-1/2}(2\pi na)\bigg],
\end{equation}
where $K_{\nu}(z)$ is the modified Bessel function of second kind and $\Gamma(z)$ is the gamma function. Using this analytic continuation, one obtains a size-independent polar
contribution plus a size-dependent contribution. We find
\begin{equation}
f_{1}(\mu)=\frac{(\mu m)^{\frac{d}{2}-1}}{4\pi^{\frac{d}{2}}}\bigg[\Gamma\left(1-\frac{d}{2}\right)+4\sum_{n=1}^{\infty}\left(2n\sqrt{\mu m}L\right)^{1-\frac{d}{2}}K_{1-\frac{d}{2}}\left(4n\sqrt{\mu m}L\right)\bigg].
\end{equation}
For odd-dimensional spaces, the gamma function is finite. For even $d$, one can use that for $n=0,1,2...$, $\Gamma(-n+\varepsilon)=\frac{(-1)^{n}}{n!}\left(\frac{1}{\varepsilon}+\psi(n+1)+\mathcal(\varepsilon)\right)$, where $\psi(z)$ is the digamma function. The introduction of bulk counterterms eliminate this divergence.

Let us discuss $f_{2}(z,\mu)$. We begin by an angular integration and $\int{}d\Omega_d=\frac{2\pi^{\frac{d}{2}}}{\Gamma(\frac{d}{2})}$ that leads to the following expression for $f_2(z,\mu)$, namely
\begin{equation}
f_2(z,\mu)=
\frac{1}{2}h(d)\int_{0}^{\infty}dp\frac{{|\p|}^{d-2}}{\sqrt{\p^2+4\mu m}} \frac{\cosh((L-2z)\sqrt{\p^2+4\mu m})}{\sinh(L\sqrt{\p^2+4\mu m})}.
\end{equation}
Using the change of variables $s=\sqrt{\p^2+4\mu m}$ in the above expression  yields the following formula for $f_{2}(z,\mu)$:
\begin{equation}
f_2(z,\mu)=\frac{1}{2}h(d)\int_{\sqrt{\mu}}^{\infty}ds(s^2-4\mu m)^{\frac{d-3}{2}}\cosh\left((L-2z)s\right)(\sinh Ls)^{-1},
\label{eq:f2-integrado-en-angulo}
\end{equation}
where $h(d)=\left[2(2{\sqrt \pi})^{d-1}\Gamma\left(\frac{d-1}{2}\right)\right]^{-1}$. 

The function $f_2(z,\mu)$ has a regular contribution plus a singular one near the boundaries. To discuss this 
singular quantity, one can show that its singular structure is the same as in the case where $\mu=0$. A complete 
analysis of these divergences are found in Refs. \cite{fo,caicedo}. Therefore we study the case where $\mu=0$.  
In fact, we are particularly interested in examining the limits ($z\rightarrow{}0^+$ and $z\rightarrow{}L^-$) which obviously contain the information about the effect of a trap. In order to fulfill this goal we introduce two new variables $x=Ls$ and $q=zs$, in terms of which we can write $f_2(z,\mu)|_{\mu=0}$ as 
\begin{equation}
f_2(z,\mu)|_{\mu=0}=\frac{h(d)}{2z^{d-2}}\int_{0}^{\infty}dq\,q^{d-3}e^{-2q}+\frac{h(d)}{2L^{d-2}}\int_{0}^{\infty}\,\, dx\,x^{d-3}
(\coth x -1)\cosh(\frac{2zx}{L}).
\label{ult2}
\end{equation}
The first term of Eq. (\ref{ult2})  gives a divergent contribution if we approach the 
boundary. The other term of Eq. (\ref{ult2})  behaves as $\frac{1}{(L-z)^{d-2}}
$. To see this let us investigate the behavior of the first integral of $f_2(z,\mu)|_{\mu=0}$ near the boundary, 
i.e., when  $z\rightarrow{}0^{+}$ and $z\rightarrow{}L^{-}$. In order to do this, we make use of two 
formulas involving the definition for the gamma function, and also another well known integral representation for the product of the gamma function times the Hurwitz zeta function given by
\begin{equation}
\int_{0}^{\infty} dx \,x^{\alpha-1}e^{-\beta x}(\coth
x-1)=2^{1-\alpha}\Gamma(\alpha)
\zeta\left(\alpha,\frac{\beta}{2}+1\right),
\label{I2}
\end{equation}
$Re(\beta)>0$, $Re(\alpha)>1$. The Hurwitz zeta function $\zeta(s,a)$ is a function of the complex variable $s=\sigma+i t$, where $\sigma, t\,\, \in{\mathbb{R}}$,  defined by the absolutely convergent series
\begin{equation}
\zeta(s,a)=\sum_{n=0}^{\infty}\frac{1}{(n+a)^{s}},\,\,\,\,Re(s)>1,
\,\,\,\,\, a \neq 0,-1,-2...
\label{na}
\end{equation}
and in the whole complex plane $\mathbb{C}$ by analytic continuation. From the definition of the gamma function and using Eq. (\ref{I2}) in Eq. (\ref{ult2}) we may write the following closed expression
\begin{align}
f_{2}(z,\mu)|_{\mu=0}&=\frac{h(d)}{2L^{d-2}}
\bigg[2^{2-d}\Gamma(d-2)\left(\zeta(d-2,\frac{z}{L}+1)+
\zeta(d-2,-\frac{z}{L}+1)\right)\bigg]\nonumber \\
&+ \frac{1}{(2z)^{d-2}}h(d)\Gamma(d-2).
\label{fim}
\end{align}
From this last expression and using the definition of the Hurwitz zeta function given by Eq. (\ref{na}), it is evident that the regularized $f_{2}(z,\mu)|_{\mu=0}$ has two poles of order $(d-2)$, one at $z=0$ and another at $z=L$. Therefore $\int_{0}^{L} dz\,f_{2}(z,\mu)|_{\mu=0}$ diverges at $z=0$ and $z=L$.  The standard way to circumvent this singular behavior is to renormalize such divergences by introducing surface counterterms at $z=0$ and $z=L$ \cite{sy}. For a careful discussion of the  physical meaning of such divergences in quantum field theory can be found in Ref. \cite{fu}. 

Hence, the zeroth-order contribution to the density of the condensate is given by
\begin{equation}
\Delta\rho^{(0)}_{R}(z,\mu)=m^{2}(f_1(\mu)-f_2(z,\mu))+ \text{bulk and surface counterterms}.
\end{equation}
There is a uniform correction to the density due to the presence of the hard walls 
associated with the  $\phi'_{1}(\x)$ field  
given by $m^{2} f_1(\mu)$ and a distortion given by $m^{2}f_2(z,\mu)$. The above calculation shows that the density profile of the condensate is highly sensitive to the presence of the hard walls potential.

\subsection{First-order corrections to the density profile}

We shall discuss the first contribution that depends on the combined effect of the trap, disorder and interactions.  The second correction to the 
density profile is $\Delta\rho^{(1)}(z_{1},\mu,\lambda_{k})$. We are going to investigate the first term in Eq. \eqref{eq:DeltaRho1}, $\Delta\rho_{1}^{(1)}(z_{1},\mu,\lambda_{k})$, which is given by 
\begin{equation}
\Delta\rho_{1}^{(1)}(z_{1},\mu,\lambda_{k})=\frac{9\,m\,\lambda_{k}}{4}\,\int\,d^{d-1}r\!\int_{0}^{L}\!\!\!\!dz\,G_{1}^{(0)}(\r,z,\r,z,\mu)G_{1}^{(0)}(\r_{1},z_{1},\r,z,\mu)G_{1}^{(0)}(\r,z,\r_{1},z_{1},\mu). \label{eq:rho11}
\end{equation}
The other contribution to $\Delta\rho^{(1)}(z_{1},\mu,\lambda_{k})$ reads:
\begin{equation}
\Delta\rho_{12}^{(1)}(z_{1},\mu,\lambda_{k})=\frac {3m}{4}\,\lambda_{k}\,\int\,d^{d-1}r\!\int_{0}^{L}\!\!\!\!dz\,G_{2}^{(0)}(\r,z,\r,z)G_{1}^{(0)}(\r_{1},z_{1},\r,z,\mu)G_{1}^{(0)}(\r,z,\r_{1},z_{1},\mu). \label{eq:rho121}
\end{equation}
The computation of these quantities are presented in details in Appendix \ref{ap:firstorder}. Both contributions show the 
presence of spatial inhomogeneities. 

\subsection{Second type of first-order contributions to the density profile}

Now, we are going to deal with the last group of first-order corrections to the density profile, 
which  contain three-point vertices. Using the Eqs. \eqref{eq:rhok}, \eqref{nova},  \eqref{nova1} and \eqref{eq:DeltaRho2}, and defining
$\Delta\rho_{1}^{(2)}(z_{1},\mu,\lambda_{k})=\Delta\rho_{1a}^{(2)}(z_{1},\mu,\lambda_{k})+\Delta\rho_{1b}^{(2)}(z_{1},\mu,\lambda_{k})$
one finds:
\begin{align}
\Delta\rho_{1a}^{(2)}(z_{1},\mu,\lambda_{k})&=\frac{9m}{2}\left(\frac{\lambda_{k}v}{2}\right)^{2}\int d^{d-1}r\int d^{d-1}r^{\prime}\int_{0}^{L}dz \int_{0}^{L}dz^{\prime}
G_{1}^{(0)}(\r_{1},z_{1},\r,z,\mu)\nonumber \\ 
&\times \left(G_{1}^{(0)}(\r,z,\r^{\prime},z^{\prime},\mu)\right)^{2}G_{1}^{(0)}(\r^{\prime},z^{\prime},\r_{1},z_{1},\mu)
\end{align}
and
\begin{align}
\Delta\rho_{1b}^{(2)}(z_{1},\mu,\lambda_{k})&=3m\left(\frac{\lambda_{k}v}{2}\right)^{2}\int d^{d-1}r\int d^{d-1}r^{\prime}\int_{0}^{L}dz \int_{0}^{L}dz^{\prime}
G_{1}^{(0)}(\r_{1},z_{1},\r,z,\mu)G_{1}^{(0)}(\r,z,\r^{\prime},z^{\prime},\mu)\nonumber \\ 
&\times G_{1}^{(0)}(\r^{\prime},z^{\prime},\r^{\prime},z^{\prime},\mu)G_{1}^{(0)}(\r,z,\r_{1},z_{1},\mu).
\end{align}
In the same way we define $\Delta\rho_{12}^{(2)}(z_{1},\mu,\lambda_{k})=\Delta\rho_{12a}^{(2)}(z_{1},\mu,\lambda_{k})+\Delta\rho_{12b}^{(2)}(z_{1},\mu,\lambda_{k})$
where $\Delta\rho_{12a}^{(2)}(z_{1},\mu,\lambda_{k})$ reads
\begin{align}
\Delta\rho_{12a}^{(2)}(z_{1},\mu,\lambda_{k})&=\frac{m}{2}\left(\frac{\lambda_{k}v}{2}\right)^{2}\int d^{d-1}r\int d^{d-1}r^{\prime}\int_{0}^{L}dz \int_{0}^{L}dz^{\prime}
G_{1}^{(0)}(\r_{1},z_{1},\r,z,\mu)\nonumber \\
&\times \left(G_{2}^{(0)}(\r,z,\r^{\prime},z^{\prime})\right)^{2}G_{1}^{(0)}(\r^{\prime},z^{\prime},\r_{1},z_{1},\mu).
\end{align}
Finally the last term $\Delta\rho_{12b}^{(2)}(z_{1},\mu,\lambda_{k})$ reads
\begin{align}
\Delta\rho_{12b}^{(2)}(z_{1},\mu,\lambda_{k})&=3m\left(\frac{\lambda_{k}v}{2}\right)^{2}\int d^{d-1}r\int d^{d-1}r^{\prime}\int_{0}^{L}dz \int_{0}^{L}dz^{\prime}
G_{1}^{(0)}(\r_{1},z_{1},\r,z,\mu)G_{1}^{(0)}(\r,z,\r^{\prime},z^{\prime},\mu)\nonumber \\ 
&\times G_{2}^{(0)}(\r^{\prime},z^{\prime},\r^{\prime},z^{\prime})G_{1}^{(0)}(\r,z,\r_{1},z_{1},\mu).
\end{align}
Inserting Eq. \eqref{geex} into the last group of first-order corrections to the density profile, one can find the contribution from $\Delta\rho_{1}^{(2)}$ and  
$\Delta\rho_{12}^{(2)}$. Also, the computation of these quantities can be obtained using the same method discussed in \ref{ap:firstorder}. However, they will not be presented here as the sums cannot be decoupled.

\section{The renormalized density of the condensate}\label{sec:finalresult}

From the above discussions, it is possible to write the renormalized density $\rho_{R}(z)$  in the case where all the terms of the series that represents the quenched free energy has effective positive coupling constants. We have
\begin{equation}
\rho_{R}(z)=C_{N}\Delta\rho^{(0)}_{R}(z,\mu)+\sum_{k=1}^{N}\frac{(-1)^{k}}{k!}\Delta\rho^{(1)}_{R}(z,\mu,\lambda_{k}),
\label{fim}
\end{equation}
where the constant $C_{N}$ is
\begin{equation}
C_{N} = \sum_{k=1}^{N}\frac{(-1)^{k}}{k!} .
\label{C_N}
\end{equation}
There are two contributions to the renormalized density. The first one depends on the presence of the hard wall. Disorder does not affect $\Delta\rho^{(0)}_{R}(z,\mu)$. This contribution is the same for the system without disorder.  It is not difficult to find the contribution of $\Delta\rho^{(0)}_{R}(\mu)$ without the trap.
The second one combines the effects of the trap, disorder and the two-body interaction, which were discussed before. These terms 
indicate the presence of inhomogeneities 
in the disordered condensate.  One can estimate the effects on the condensate, due to 
disorder, as follows: the case without disorder is given by $\lambda_{k}=g=\text{constant}$, so $\Delta
\rho^{(1)}_{R}(z,\mu,\lambda_{k})=\Delta\rho^{(1)}_{R}(z,\mu,g)$, which does not depend on $k$. The difference of the renormalized density with and without disorder is
\begin{equation}
\sum_{k=1}^{N}\frac{(-1)^{k}}{k!}\Delta\rho^{(1)}_{R}(z,\mu,\lambda_{k})-C_{N}\Delta\rho^{(1)}_{R}(z,\mu,g).
\end{equation}

In the same way that we obtained from the disordered-average generating functional of connected correlation functions the renormalized density $\rho_{R}(z)$ in the first-order 
approximation of the effective coupling constant, it is possible to get the renormalized coupling constant between bosonic particles also in this first-order approximation. For attractive boson-boson interactions, 
the disorder produces  an effect similar to that that of a Feshbach resonance, in
that it changes the sign of the two-body interaction between the atoms~\cite{chin}. Therefore, the  scattering 
length becomes negative due to the effects of the disorder in this case.

For a weak disorder, one can increase the number of particles to obtain a negative effective coupling constant. Or one can also obtain a negative coupling constant increasing the strength of the disorder.  The next problem which arises in the case where the disorder is strong.  There is a critical $k$ such that $k_{c}=\left\lfloor\frac{g}{\beta\sigma}\right\rfloor$.  In this case the quenched free energy has two contributions. The first one, for $k\leq k_{c}$ was analyzed. In the second one, the effective coupling constants are negative. In the conclusions we discuss briefly this situation.


\section{Conclusions and Outlook}\label{sec:conclusions}

In this work  we study the effects of disorder in a dilute Bose-Einstein 
condensate confined in a hard walls trap. We analyze the behavior of the system in equilibrium at 
temperatures below the critical condensation temperature for different values of the strength of the disorder. Physical quantities 
are calculated from the quenched free energy. 

We start working at the mean-field level,  where the random system is described by the disordered 
Gross-Pitaevskii  energy functional. Then, the quenched  free energy 
is written in a series of moments of the partition function of the system.  This series 
representation describes the system as an ensemble of subsystems, each of them being characterized by
 a disorder-dependent effective coupling constant that can be positive or negative.
In a generic moment of the partition function, the effective coupling constant becomes negative 
for some critical value of the strength of the disorder. We have implemented an analysis for a generic integer moment, in which the system is 
described by $k$ identical and non-interacting complex fields. Since all the fields are equal, one can 
study the contribution  from  a single field. The ground state of such system is discussed for different 
strengths of the disorder. 

In the first-order approximation of the effective coupling constant, we discuss the combined effect of the hard walls and the two-body interaction.  This can be done by studying the behavior of the two-point static correlation functions. Studying a generic integer moment with  positive effective 
 coupling constant,  which corresponds to  weak disorder, we present a general expression for its contribution to the density profile of the condensate. There are two contributions to the renormalized density. The first one depends on the presence of the hard walls. The second one combines effects of the trap, disorder and the two-body interaction. 

Since it is possible to control the strength of the disorder  thereby changing  
the magnitude of the effective coupling constants, it is always possible to produce a 
scenario for effective negative coupling constant in some integer moment of the partition function. 
In such a situation,  it is necessary to introduce  a three-particle interaction contribution~\cite{hugenholz} 
for defining a potential $\Phi(\phi_{1},\phi_{2})$ with a well-defined extremum.   
The new effective Gross-Pitaevskii energy functional, $E^{\prime\prime}_{\text{eff}}(\phi_{1},\phi_{2})$, 
can be written as 
\begin{equation}
E^{\prime\prime}_{\text{eff}}(\phi_{1},\phi_{2})=E^{\prime}_{\text{eff}}(\phi_{1},\phi_{2})+\int d^{d}x\;  \xi\; (\phi_{1}^{2}+\phi_{2}^{2})^{3},
\end{equation}
where $E^{\prime}_{\text{eff}}(\phi_{1},\phi_{2})$ is given by Eq. \eqref{eq:effectivehamiltoniantt}. Note that  
we could have anticipated the need of such a term and included it from the very beginning in the Hamiltonian of  
Eq.~(\ref{eq:hamiltonian}), as one would do within an effective field theory framework~\cite{braaten}.  
Bose-Einstein condensate with a three-particle interaction term was also discussed in Ref. 
\cite{shukla, gamall}.  The analysis of the existence of first and second order phase transitions for such 
functional with three-particle interaction contribution were discussed in Ref. \cite{binder}. This first-order 
phase transition in the condensate is induced by disorder fields. 

A natural continuation of this paper is to use the series representation for the quenched free energy to discuss 
the behavior of the Bose-Einstein condensate with disorder in three different situations. First, still in the 
scenario of a stationary situation, discuss the strong disorder limit by taking into 
account the second  contribution to the series defined by Eq. \eqref{eq:completefreeenergyseries}, 
 and find the ground state of 
the system for strong-disorder fields. Second, going back to the weak disorder limit introduce dynamics in the 
problem, i.e., using 
 Eq. \eqref{decomp}, instead of the static Gross-Pitaevskii equation. In this second situation one is 
interested to show, for instance how the disorder modifies the sound velocity in the condensate.
Finally the strong disorder limit with dynamics must be analyzed. These subjects are under investigation by the 
authors.

\section*{Acknowledgments}

We would like to thank G. Menezes and for useful discussions. This work was partially suported by Conselho Nacional de Desenvolvimento Cient\'{\i}fico e Tecnol\'{o}gico - CNPq, 309982/2018-9 (C.A.D.Z.), 305894/2009-9 (G.K.), 303436/2015-8 (N.F.S.), INCT F\'{\i}sica Nuclear e Apli\-ca\-\c{c}\~oes, 464898/2014-5  (G.K) and Funda\c{c}\~{a}o de Amparo \`{a} Pesquisa do Estado de S\~{a}o Paulo - FAPESP, 2013/01907-0 (G.K).

\appendix

\section{Computations of $\Delta\rho^{1}$}\label{ap:firstorder}

In this Appendix, we will present a detailed derivation of $\Delta\rho^{1}$. From Eqs.  \eqref{MF} and\eqref{eq:DeltaRho2} one finds
\begin{equation}
\Delta\rho_{1}^{(1)}=\frac{m^{2}\lambda_{k}}{2}\int\,d^{d-1}r\!\int_{0}^{L}\!\!\!\!dz
\,G_{1}^{(0)}(\r_{1},z_{1},\r,z,\mu) G_{1}^{(0)}(\r,z,\r,z,\mu)
G_{1}^{(0)}(\r,z,\r_{1},z_{1},\mu).
\end{equation}
Inserting Eq. \eqref{geex} into the above equation, one has
\begin{align}
\Delta\rho_{1}^{(1)}&=\left(\frac{2}{L}\right)^{3}\frac{m^{5}\lambda_{k}}{2(2\pi)^{3(d-1)}}\sum_{j,k,l=1}^{\infty}\int\!d^{d-1}r\!\int_{0}^{L}\!\!\!\!dz\!\int\!d^{d-1}p_{1}\!\int\!d^{d-1}p_{2}\!\int\!d^{d-1}p_{3}\nonumber \\
&\times\sin\left(\frac{j\pi z_{1}}{L}\right)\sin\left(\frac{j\pi z}{L}\right)\sin^{2}\left(\frac{k\pi z}{L}\right)\sin\left(\frac{l\pi z}{L}\right)\sin\left(\frac{l\pi z_{1}}{L}\right)\nn \\
&\times\frac{e^{i\p_{1}\cdot(\r_{1}-\r)}e^{-i\p_{3}\cdot(\r_{1}-\r)}}{\left(\p_{1}^{2}+\left(\frac{j\pi}{L}\right)^{2}+4\mu m\right)\left(\p_{2}^{2}+\left(\frac{k\pi}{L}\right)^{2}+4\mu m\right)\left(\p_{3}^{2}+\left(\frac{l\pi}{L}\right)^{2}+4\mu m\right)} \label{eq:deltarho1completo}
\end{align}
The integral in $\r$ can be performed using the relation
\begin{equation}
\int d^{d-1}r  e^{i(\p_{1}-\p_{3})\cdot\r}=(2\pi)^{d-1}\delta^{d-1}(\p_{1}-\p_{3}).
\end{equation}
From the above equation the integral in $\p_{3}$ can be also evaluated. It remains to compute the integral in $z$. The product of sines can be written as:
\begin{align}
&\sin\left(\frac{j\pi z}{L}\right)\sin^{2}\left(\frac{k\pi z}{L}\right)\sin\left(\frac{l\pi z}{L}\right)=\sin\left(\frac{j\pi z}{L}\right)\sin\left(\frac{l\pi z}{L}\right)\nonumber \\
&-\frac{1}{4}\left[\sin\left(\frac{(j+k)\pi z}{L}\right)\sin\left(\frac{(l+z)\pi z}{L}\right)\sin\left(\frac{(j-k)\pi z}{L}\right)\sin\left(\frac{(l-k)\pi z}{L}\right)\right.\nonumber \\
&\left.-\sin\left(\frac{(j+k)\pi z}{L}\right)\sin\left(\frac{(l-k)\pi z}{L}\right)-\sin\left(\frac{(j-k)\pi z}{L}\right)\sin\left(\frac{(l+k)\pi z}{L}\right)\right].
\end{align}
The $z$ integral can be computed using the completeness of $u_{n}(z)$:
\begin{equation}
\int_{0}^{L}dz \sin\left(\frac{m\pi z}{L}\right)\sin\left(\frac{n\pi z}{L}\right)=\frac{L}{2}\delta_{mn}.
\end{equation}
Thefore Eq. \eqref{eq:deltarho1completo} can be written as
\begin{equation}
\Delta\rho_{1}^{(1)}=\Delta\rho_{1\textbf{(a)}}^{(1)}+\Delta\rho_{1\textbf{(b)}}^{(1)}, 
\end{equation}
where
\begin{align}
\Delta\rho_{1\textbf{(a)}}^{(1)}&=\left(\frac{2}{L}\right)^{2}\frac{m^{5}\lambda_{k}}{4(2\pi)^{2(d-1)}}\left[\sum_{j=1}^{\infty}\sin^{2}\left(\frac{j\pi z_{1}}{L}\right)\int\!d^{d-1}p_{1}\frac{1}{\left(\p_{1}^{2}+\left(\frac{j\pi}{L}\right)^{2}+4\mu m\right)^{2}}\right]\nonumber  \\ 
&\times\left[\sum_{k=1}^{\infty}\int\!d^{d-1}p_{2}\frac{1}{\left(\p_{2}^{2}+\left(\frac{k\pi}{L}\right)^{2}+4\mu m\right)^{2}}\right]
\label{eq:deltarho1a}
\end{align}
and
\begin{align}
&\Delta\rho_{1\textbf{(b)}}^{(1)}=\frac{m^{5}\lambda_{k}}{(2\pi)^{2(d-1)}L^{2}}\sum_{j,l=1}^{\infty}\sin\left(\frac{j\pi z_{1}}{L}\right)\sin\left(\frac{l\pi z_{1}}{L}\right)\nonumber \\
&\times\int\!d^{d-1}p_{1}\frac{1}{\left(\p_{1}^{2}+\left(\frac{j\pi}{L}\right)^{2}+4\mu m\right)\left(\p_{1}^{2}+\left(\frac{l\pi}{L}\right)^{2}+4\mu m\right)}\int\!d^{d-1}p_{2}\frac{1}{\left(\p_{2}^{2}+\left(\frac{(l-j)\pi}{L}\right)^{2}+4\mu m\right)^{2}}.
\label{eq:deltarho1b}
\end{align}

The quantity $\Delta\rho_{12}^{(1)}$  is given by
\begin{equation}
\Delta\rho_{12}^{(1)}=\frac{m^{2}\lambda_{k}}{2}\int\,d^{d-1}r\!\int_{0}^{L}\!\!\!\!dz
\,G_{1}^{(0)}(\r_{1},z_{1},\r,z,\mu)G_{2}^{(0)}(\r,z,\r,z)
G_{1}^{(0)}(\r,z,\r_{1},z_{1},\mu).
\end{equation}
This quantity can be computed by the same steps done to calculate $\Delta\rho_{1}^{(1)}$ as the unique change consists in taking the limit $\mu\to 0$ in the $\p_{2}$ integral. 

In order to renormalize $\Delta\rho^{(1)}_{\textbf{(a)}}$, we will follow the same steps as before to renormalize $\Delta\rho^{(0)}$. There are contributions similar to $f_{1}(\mu)$ and $f_{2}(\mu,z)$, however the complete analysis of that terms due to its complexity will be omitted here. The second term, $\Delta\rho^{(1)}_{\textbf{(b)}}$ cannot be computed directly as the sums are not decoupled. However, one can integrate in $z_{1}$ to circumvented this problem. Again, we will omit that computation. The same considerations apply to $\Delta\rho_{12\textbf{(a)}}^{(1)}$ and $\Delta\rho_{12\textbf{(b)}}^{(1)}$. These quantities can be written as
\begin{equation}
\Delta\rho_{12}^{(1)}=\Delta\rho_{12\textbf{(a)}}^{(1)}+\Delta\rho_{12\textbf{(b)}}^{(1)}, 
\end{equation}
where $\Delta\rho_{12\textbf{(a)}}^{(1)}$ is given by
\begin{align}
\Delta\rho_{12\textbf{(a)}}^{(1)}&=\left(\frac{2}{L}\right)^{2}\frac{m^{5}\lambda_{k}}{4(2\pi)^{2(d-1)}}\left[\sum_{j=1}^{\infty}\sin^{2}\left(\frac{j\pi z_{1}}{L}\right)\int\!d^{d-1}p_{1}\frac{1}{\left(\p_{1}^{2}+\left(\frac{j\pi}{L}\right)^{2}+4\mu m\right)^{2}}\right]\nonumber  \\ 
&\times\left[\sum_{k=1}^{\infty}\int\!d^{d-1}p_{2}\frac{1}{\left(\p_{2}^{2}+\left(\frac{k\pi}{L}\right)^{2}\right)^{2}}\right]
\label{eq:deltarho12a}
\end{align}
and $\Delta\rho_{12\textbf{(b)}}^{(1)}$ is
\begin{align}
&\Delta\rho_{12\textbf{(b)}}^{(1)}=\frac{m^{5}\lambda_{k}}{(2\pi)^{2(d-1)}L^{2}}\sum_{j,l=1}^{\infty}\sin\left(\frac{j\pi z_{1}}{L}\right)\sin\left(\frac{l\pi z_{1}}{L}\right)\nonumber \\
&\times\int\!d^{d-1}p_{1}\frac{1}{\left(\p_{1}^{2}+\left(\frac{j\pi}{L}\right)^{2}+4\mu m\right)\left(\p_{1}^{2}+\left(\frac{l\pi}{L}\right)^{2}+4\mu m\right)}\int\!d^{d-1}p_{2}\frac{1}{\left(\p_{2}^{2}+\left(\frac{(l-j)\pi}{L}\right)^{2}\right)^{2}}.
\label{eq:deltarho12b}
\end{align}
%

\end{document}